\documentclass{emulateapj}
\usepackage{apjfonts}
\begin{document}
\title{Contribution of the Gamma-ray Loud Radio Galaxies Core Emissions to \\ the Cosmic MeV and GeV Gamma-Ray Background Radiation}
\author{Yoshiyuki Inoue$^{1,2}$}
\affil{$^1$Department of Astronomy, Kyoto University, Kitashirakawa, Sakyo-ku, Kyoto 606-8502, Japan}
\affil{$^2$Max Planck Institute for Physics, F${\rm \ddot{o}}$hringer Ring 6, 80805 Munich, Germany}
\email{yinoue@kusastro.kyoto-u.ac.jp}

\begin{abstract} 
The {\it Fermi} gamma-ray satellite has recently detected gamma-ray emissions from radio galaxy cores. From these samples, we first examine the correlation between the luminosities at 5 GHz, $L_{\rm 5GHz}$, and at 0.1-10 GeV, $L_{\gamma}$, of these gamma-ray loud radio galaxies. We find that the correlation is significant with $L_\gamma\propto L_{\rm 5GHz}^{1.16}$ based on a partial correlation analysis. Using this correlation and the radio luminosity function (RLF) of radio galaxies, we further explore the contribution of  gamma-ray loud radio galaxies to the unresolved extragalactic gamma-ray background (EGRB). The gamma-ray luminosity function is obtained by normalizing the RLF to reproduce the source count distribution of the {\it Fermi} gamma-ray loud radio galaxies. We find that gamma-ray loud radio galaxies will explain $\sim25$\% of the unresolved {\it Fermi} EGRB flux above 100 MeV and will also make a significant contribution to the EGRB in the 1-30 MeV energy band. Since blazars explain 22\% of the EGRB above 100 MeV, radio loud active galactic nuclei (AGNs) populations explain $\sim47$\% of the unresolved EGRB. We further make an interpretation on the origin of the EGRB. The observed EGRB spectrum at 0.2-100 GeV does not show an absorption signature by the extragalactic background light. Thus, the dominant population of the origin of EGRB at very high energy ($>30$GeV) might be nearby gamma-ray emitting sources or sources with very hard gamma-ray spectrum.
\end{abstract}

\keywords{cosmology: diffuse radiation -- galaxies : active -- gamma rays : theory}

\section{Introduction}
\label{intro}
The origin of the extragalactic diffuse MeV and GeV gamma-ray background (EGRB) radiation has been argued for a long time in astrophysics, although it is well known that radio quiet active galactic nuclei (AGNs) take account for the cosmic X-ray background (CXB) below several hundred keV \citep[for reviews][]{bol87,fab92,ued03,has05,gil07}. The EGRB spectrum at 0.3--30 MeV is measured by {\it SMM} \citep{wat97} and COMPTEL on board the Compton Gamma-Ray Observatory \citep{kap96}. In the GeV energy range, EGRB was first discovered by the {\it SAS}--2 satellite \citep{fic78,tho82}. EGRET (Energetic Gamma-Ray Experiment Telescope) on board the Compton Gamma-Ray Observatory confirmed the EGRB spectrum at 0.03-50 GeV \citep{sre98,str04}. Recently,  LAT (Large Area Telescope) on board the {\it Fermi} Gamma-ray Space Telescope ({\it Fermi}) made a new measurement of the EGRB spectrum from 0.2 to 100 GeV \citep{abd10_egrb}. The observed integrated EGRB flux ($E>100$ MeV) is $1.03\times10^{-5}$ photons/cm$^2$/s/sr with a photon index of 2.41.

Several sources have been suggested to explain the MeV background. One is the nuclear decay gamma-ray from Type Ia supernovae \citep[SNe Ia][]{cla75,zdz96,wat99}. However, the recent measurement of the cosmic SNe Ia rate suggests that the expected flux from SNe Ia is one order of magnitude lower than the measured MeV EGRB flux \citep{ahn05_SN,str05,hor10}. Comptonization emission from non-thermal electrons in AGN coronae is also proposed \citep{ino08}. This model explains the origin of CXB and the MeV EGRB in the same population. Blazars, which is one type of AGNs with the direction of a relativistic jet coinciding with our line of sight, are also proposed \citep{aje09}. Very recently, \citet{mas11} has shown that the gamma-ray emission from lobes of radio galaxies will explain $\sim10$\% of the MeV background flux. MeV mass scale dark matter (DM) annihilations has also been discussed \citep{ahn05a,ahn05b}, but there is no natural particle physics candidate for a dark matter with a mass scale of MeV energies. However it is still discussing about the origin of the MeV background due to the difficulties of MeV gamma-ray measurements. 

In the case of the GeV background, since blazars are dominant extragalactic gamma-ray sources \citep{har99,abd10_agn}, it is expected that unresolved population of blazars would explain the GeV EGRB \citep{pad93,ste93,sal94,chi95,ste96,chi98,muk99,muc00,nar06,gio06,der07,pav08,kne08,bha09,ino09}. Recently, \citet{abd10_marco} showed that unresolved blazars can explain $\sim22$\% of the EGRB above 0.1 GeV by analyzing 11-month {\it Fermi} AGN catalog. Very recently \citet{ste10} proposed that the unresolved blazar population would be able to explain the EGRB spectrum below 1 GeV by taking into account of the energy dependence of source confusion effects. 

Other gamma-ray emitting extragalactic sources have also been discussed as the origin of the GeV EGRB. Those are intergalactic shocks produced by the large scale structure formation \citep{loe00,tot00,min02,kes03,gab03}, normal and starburst galaxies \citep{pav02,tho07,bha09_gal,mak10,fie10}, high Galactic latitude pulsars \citep{fau10,sie10}, kilo-parsec (kpc) size AGN jets \citep{sta06}, radio quiet AGNs \citep{ino08,ino09}, and GeV mass scale DM annihilation or decay \citep[see e.g.][]{jun96,ber00,uli02,oda05,and06,hor06,and07,ahn07,and09,kaw09}.

Here, {\it Fermi} has recently detected GeV gamma-ray emissions from 11 misaligned AGNs (i.e. radio galaxies), which are one type of AGNs with the direction of a relativistic jet {\it not} coinciding with our line of sight \citep{abd10_core}. Their average photon index at 0.1-10 GeV is $\sim2.4$ which is same as that of GeV EGRB \citep{abd10_egrb} and blazars \citep{abd10_marco}. Although they are fainter than blazars, the expected number in the entire sky is much more than blazars. Then, it is naturally expected that they will make a significant contribution to EGRB. Therefore, in this paper, we study the contribution of such gamma-ray loud radio galaxies (not blazars) to EGRB. 

To study the EGRB contribution of gamma-ray loud radio galaxies, their gamma-ray luminosity function (GLF) is required. Because of limited samples, it is not a straightforward task to construct it using only {\it Fermi} samples. Therefore, we first investigate the correlation between radio and gamma-ray luminosities of gamma-ray loud radio galaxies. In the case of blazars, the correlation between gamma-ray and radio luminosities have been presented in many papers since EGRET era \citep{pad93,ste93,sal94,don95,zha01,nar06,ghi10a,ghi10b}. Since the radio luminosity function (RLF) of radio galaxies is well studied \citep[see e.g.][]{dun90,wil01}, we are able to obtain the GLF by converting the RLF to the GLF using a luminosity correlation. With that GLF, we evaluate their contribution to EGRB.

This work is organized as follows. Samples used in this study are shown in \S.\ref{sec:sam}. In \S.\ref{sec:glf}, we will investigate the radio and gamma-ray luminosity correlation and determine our GLF to reproduce the number of detected gamma-ray loud radio galaxies. The EGRB will be calculated and compared with the observed data in \S.\ref{sec:egrb}. Discussion and conclusion will be given in \S.\ref{sec:dis} and \S.\ref{sec:con}, respectively. Throughout this paper, we adopt the standard cosmological parameters of $(h,\Omega_M,\Omega_\Lambda) = (0.7, 0.3, 0.7)$.

\section{Samples}
\label{sec:sam}

{\it Fermi} has reported the detection of 11 Fanarof-Riley (FR) radio galaxies including 7 type-I FR (FRI) galaxies and 4 type-II FR (FRII) galaxies by the entire sky survey for 15 months \citep{abd10_core}. FRI galaxies have decelerating jets, knots at kpc distance from the core and edge-darkened lobes, while FRII galaxies have relativistic jets and edge-brightened radio lobes with bright hotspots \citep{fan74}. In the scheme of the AGN jet unification scenario, FRI and FRII galaxies are the misaligned AGN populations of BL Lac objects and flat spectrum radio quasars (FSRQs), respectively \citep{urr95}. 

In this study, we use 10 samples (6 FRIs and 4 FRIIs) from \citet{abd10_core} which were already reported in 11-month {\it Fermi} catalog \citep{abd10_catalog,abd10_agn}. This is because, as discussed in \S.\ref{sec:glf} below, we use the detection efficiency of {\it Fermi} shown in \citet{abd10_marco} which are constructed from the 11-month catalog. Table. 1 lists the gamma-ray loud radio galaxy samples used in this study. It gives the object name, the First Source Catalog (1FGL) {\it Fermi}-LAT source name, redshift, gamma-ray photon flux above 0.1 GeV, photon index at 0.1-10 GeV,  radio flux at 5 GHz, spectral index at 5 GHz, radio classification. The definition of photon index, $\Gamma$, and spectral index, $\alpha$, is the index of power-law differential photon spectrum, $dN/d\epsilon\propto\epsilon^{-\Gamma}$ , and $\Gamma-1$, respectively. 

\begin{deluxetable*}{llrrrrrcc}[h]
\tabletypesize{\scriptsize}
\tablecaption{Observed parameters of gamma-ray loud radio galaxies.
\label{tb:sam}}
\tablewidth{0pt}
\tablehead{
\colhead{Object Name} 
& \colhead{1FGL Name}
& \colhead{$z$}
& \colhead{$F_\gamma$}
& \colhead{$\Gamma$}  
& \colhead{$S_{\rm R}$}
& \colhead{$\alpha_{r}$}
& \colhead{Class}
& \colhead{REF}\\
&  & & [$\times10^{-9}$ ph/cm$^2$/s] & &  [Jy] & }
\startdata
3C 78/NGC 1218   &1FGLJ 0308.3+0403& 0.029   & $4.7\pm1.8$ &  $1.95\pm0.14$ &$0.964\pm0.048$ &0.64& FRI&1\\ 
3C 84/NGC 1275   &1FGLJ 0319.7+4130& 0.018   & $222\pm8$ &  $2.13\pm0.02$ &$3.10\pm0.02$ &0.78&FRI&2\\ 
3C 111   			 &1FGLJ 0419.0+3811& 0.049   & $40\pm8$ &  $2.54\pm0.19$&$1.14\pm0.0^{a}$ &-0.146&FRII&3\\ 
PKS 0625-354   	&1FGLJ 0627.3-3530& 0.055   & $4.8\pm1.1$ & $2.06\pm0.16$ &$0.60\pm0.03$ &0.53&FRI&1\\ 
3C 207 			&1FGLJ 0840.8+1310& 0.681   & $24\pm4$ &  $2.42\pm0.10$ &$0.51\pm0.02$ &0.9&FRII&2\\ 
PKS 0943-76 		&1FGLJ 0940.2-7605& 0.27   & $55\pm12$ &  $2.83\pm0.16$ &$0.79\pm0.03$ &0.79&FRII&4\\ 
M87/3C 274 		&1FGLJ 1230.8+1223& 0.004   & $24\pm6$ &  $2.21\pm0.14$ &$4.0\pm0.04$ &0.79&FRI&2\\ 
Cen A   			&1FGLJ 1325.6-4300& 0.0009   & $214\pm12$ &  $2.75\pm0.04$ &$6.98\pm0.21$ &1.2&FRI&1\\ 
NGC 6251  		&1FGLJ 1635.4+8228& 0.024   & $36\pm8$ &  $2.52\pm0.12$ &$0.35\pm0.045$ &0.72 &FRI&2\\ 
3C 380   			 &1FGLJ 1829.8+4845& 0.0692   & $31\pm18$ &  $2.51\pm0.30$ &$7.45\pm0.047$ & 0.71 &FRII& 2
\enddata
\tablecomments{ 
1FGL Name: the First Source Catalog (1FGL) {\it Fermi}-LAT source name,
$z$: redshift of the source,
$F_\gamma$: gamma-ray photon flux above 100 MeV in $10^{-9}$ photons/cm$^2$/s,  
$\Gamma$: photon index at 0.1-10 GeV,  
$S_{\rm R}$: radio flux density at 5\,GHz in Jy,  
$\alpha_{r}$: radio spectral index at 5\,GHz,  
Class: FRI is type I of Fanaroff-Riley galaxy and FRII is type II of Fanaroff-Riley galaxy \citep{fan74}.
References --- (1) \citet{ung84,sai86,bau88,eke89,jon92,bur93,mor93}; (2) \citet{ben62,spi85,lai83}; (3) \citet{lin84}; (4) \citet{bur06a,bur06b}.}
\tablenotetext{a}{No error is reported in \citet{lin84}.}
\end{deluxetable*}

From Table. 1, the mean photon index of gamma-ray spectra at 0.1--10 GeV, $\Gamma_c$, is 2.39 and the spread is 0.28. This  index is same as that of blazars, 2.40 \citep{abd10_marco}, and the EGRB spectrum, 2.41 \citep{abd10_egrb}.

From the individual source studies \citep{abd09_ngc1275,abd09_m87,abd10_cena}, the typical gamma-ray SEDs are well explained by synchrotron-self-Compton emission models. Here, by taking account of particle cooling effect, the electron spectrum is given by  $dN/d\gamma_e \propto \gamma_e^{-p}$ at $\gamma_e \le \gamma_{\rm br}$ and $dN/d\gamma_e \propto \gamma_e^{-(p+1)}$ at $\gamma_e > \gamma_{\rm br}$, where $\gamma_{\rm br}$ is the cooling break lorentz factor. The inverse Compton (IC) photon spectrum is given by 
\begin{eqnarray}  
  dN/d\epsilon\propto\left\{\begin{array}{ll}
     \epsilon^{-(p+1)/2} & \ \ \epsilon \le \epsilon_{\rm br}, \\
       \epsilon^{-(p+2)/2} & \ \ \epsilon > \epsilon_{\rm br}, \\
    \end{array}\right.
    \label{eq:sed}
\end{eqnarray} 
where $\epsilon_{\rm br}$ corresponds to the IC photon energy from electrons with $\gamma_{\rm br}$ \citep{ryb79}.

 The SED fitting for NGC 1275 and M87 show that the IC peak energy in the rest frame locates at $\sim5$ MeV \citep{abd09_ngc1275,abd09_m87}. In this study, we use the mean photon index, $\Gamma_c$, as $\Gamma$ at 0.1--10 GeV and we set a peak energy, $\epsilon_{\rm br}$, in photon spectrum at 5 MeV for all gamma-ray loud radio galaxies as a baseline model. Then, we are able to define the average SED shape of gamma-ray loud radio galaxies for all luminosities as $dN/d\epsilon\propto\epsilon^{-2.39}$ at $\epsilon>$5 MeV, and $dN/d\epsilon\propto\epsilon^{-1.89}$ at $\epsilon\le$5 MeV by following Equation \ref{eq:sed}.

However, only 3 sources are currently studied with multi-wavelength observational data. We need to make further studies of individual gamma-ray loud radio galaxies to understand their SED properties in wide luminosity ranges. We examine other spectral models in the Section \ref{subsec:unc}.

\section{Gamma-ray Luminosity Function}
\label{sec:glf}

\subsection{Radio and Gamma-ray Luminosity Correlation}
\label{subsec:lrlg}

To estimate the EGRB contribution of gamma-ray loud radio galaxies, we need to construct a GLF. However, because of a small sample size, it is difficult to construct a GLF using the current gamma-ray data only. Here, the RLF of radio galaxies is extensively studied by previous works \citep[see e.g.][]{dun90,wil01}. If there is a correlation between the radio and gamma-ray luminosities, we are able to convert the RLF to the GLF with that correlation. In the case of blazars, it has been suggested that there is a correlation between radio and gamma-ray luminosity from the EGRET era \citep{pad93,ste93,sal94,don95,zha01,nar06}, although it has also been discussed that this correlation can not be firmly established because of flux limited samples \citep{muc97}. Recently, using the {\it Fermi} samples, \citet{ghi10a,ghi10b} confirmed that there is a correlation between the radio and gamma-ray luminosities.

To examine a luminosity correlation in gamma-ray loud radio galaxies, we first derive the radio and gamma-ray luminosity of gamma-ray loud radio galaxies as follows. Gamma-ray luminosities between the energies $\epsilon_1$ and $\epsilon_2$ are calculated by
\begin{equation}
L_\gamma(\epsilon_1,\epsilon_2) = 4\pi d_L(z)^2 \frac{S_\gamma(\epsilon_1,\epsilon_2)}{(1+z)^{2-\Gamma}},
\end{equation}
where $d_L(z)$ is the luminosity distance at redshift, $z$, $\Gamma$ is the photon index and $S(\epsilon_1,\epsilon_2)$ is the observed energy flux between the energies $\epsilon_1$ and $\epsilon_2$. The energy flux is given from the photon flux $F_\gamma$, which is in the unit of photons/cm$^2$/s, above $\epsilon_1$ by
\begin{eqnarray}
S_\gamma(\epsilon_1,\epsilon_2) &=& \frac{(\Gamma-1)\epsilon_1}{2-\Gamma} \left[\left(\frac{\epsilon_2}{\epsilon_1}\right)^{2-\Gamma}-1\right] F_\gamma, \ (\Gamma\neq2)\\
S_\gamma(\epsilon_1,\epsilon_2) &=& \epsilon_1\ln({\epsilon_2}/{\epsilon_1})  F_\gamma,\ \ \ \ \ (\Gamma=2).
\end{eqnarray}
Radio luminosity is also calculated in the same manner.

\begin{figure}
  \begin{center}
\centering
\plotone{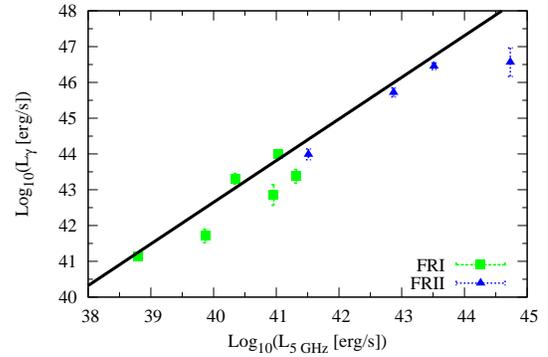}
\caption{Gamma-ray luminosity at 0.1--10 GeV versus radio luminosity at 5 GHz. The square and triangle data represents FRI and FRII galaxies, respectively. The solid line is the fit to all sources.}
\label{fig:lrlg}
\end{center}
\end{figure}

Figure \ref{fig:lrlg} shows the 5 GHz and 0.1-10 GeV luminosity relation of {\it Fermi} gamma-ray loud radio galaxies. Square and triangle data represents FRI and FRII radio galaxies, respectively. The solid line shows the fitting line to all the data. The function is given by
\begin{equation}
\log_{10} (L_\gamma) = (-3.90\pm0.61) + (1.16\pm0.02)\log_{10}(L_{\rm 5 GHz}),
\label{eq:lrlg}
\end{equation}
where errors show 1-$\sigma$ uncertainties. In the case of blazars, the slope of the correlation between $L_\gamma(>100 {\rm MeV})$, luminosity above 100 MeV, and radio luminosity at 20 GHz is $1.07\pm0.05$ \citep{ghi10b}. The correlation slopes of gamma-ray loud radio galaxies are similar to that of blazars. This may indicate that emission mechanism is similar in gamma-ray loud radio galaxies and blazars.

We need to examine whether the correlation between the radio and gamma-ray luminosities is true or not. In the flux limited observations, the luminosities of samples are strongly correlated with redshifts. This might result in a spurious luminosity correlation. As in previous works on blazar samples \citep{pad92,zha01,ghi10b}, we perform a partial correlation analysis to test the correlation between the radio and gamma-ray luminosities excluding the redshift dependence (see the Appendix for details). First, we calculate the Spearman rank-order correlation coefficients \citep[see e.g.][]{nr92}. The correlation coefficient is 0.993, 0.993, 0.979 between $\log_{10}L_{\rm 5 GHz}$ and $\log_{10}L_\gamma$, between $\log_{10}L_{\rm 5 GHz}$ and redshift, and between $\log_{10}L_\gamma$ and redshift, respectively. Then, the partial correlation coefficient becomes 0.866 with chance probability $1.65\times10^{-6}$. Therefore, we conclude that there is a correlation between the radio and gamma-ray luminosities of gamma-ray loud radio galaxies.

\subsection{Gamma-ray Luminosity Function}
\label{subsec:glf}

In this section, we derive the GLF of gamma-ray loud radio galaxies, $\rho_\gamma(L_\gamma,z)$. There is a correlation between the radio and gamma-ray luminosities as Equation \ref{eq:lrlg}. With this correlation, we develop the GLF by using the RLF of radio galaxies, $\rho_r(L_r,z)$, with radio luminosity, $L_r$. The GLF is given as 
\begin{equation}
\rho_\gamma(L_\gamma,z) = \kappa \frac{dL_r}{dL_\gamma}\rho_r(L_r,z),
\label{eq:lf}
\end{equation}
where $\kappa$ is a normalization factor. We use the 151 MHz RLF \citep{wil01}. Since they presented the formula of FRI and FRII RLFs separately, we combined them as in their paper because it is difficult to analyze each population separately with our limited number of samples. Moreover, since the cosmological parameters in \citet{wil01} are $\Omega_M=\Omega_\Lambda=0$ and $h=0.5$, we also convert the RLF to the standard cosmology adopted in this study. The RLF is given by 
\begin{equation}
\rho_r(L_r,z) = \eta(z) \times [\rho_{r,{\rm FRI}}(L_r,z) + \rho_{r,{\rm FRII}}(L_r,z)],
\end{equation}
where $\eta(z)$ is the conversion factor of cosmology, $\rho_{r,{\rm FRI}}$ is the FRI RLF, and $\rho_{r,{\rm FRII}}$ is the FRII RLF. The FRI RLF and FRII RLF are given by 
\begin{eqnarray}  
  \rho_{r,{\rm FRI}}(L_r,z)=\left\{\begin{array}{ll}
      \rho_{\rm I,0}\left(\frac{L_r}{L_{\rm I,c}}\right)^{-\alpha_{\rm I}} \exp\left(\frac{-L_r}{L_{\rm I,c}}\right)(1+z)^{k_{\rm I}} & z \le z_{\rm I,c}, \\
       \rho_{\rm I,0}\left(\frac{L_r}{L_{\rm I,c}}\right)^{-\alpha_{\rm I}} \exp\left(\frac{-L_r}{L_{\rm I,c}}\right)(1+z_{\rm I,c})^{k_{\rm I}} & z > z_{\rm I,c}, \\
    \end{array}\right.
\end{eqnarray} 

\begin{equation}
\rho_{r,{\rm FRII}}(L_r,z)=\rho_{\rm II,0}\left(\frac{L_r}{L_{\rm II,c}}\right)^{-\alpha_{\rm II}} \exp\left(\frac{-L_{\rm II,c}}{L_r}\right)f_{\rm II}(z).
\end{equation}
Here, $f_{\rm II}(z)$ is the evolution function and given by
\begin{eqnarray}  
  f_{\rm II}(z)=\left\{\begin{array}{ll}
      \exp\left[-\frac{1}{2}\left(\frac{z-z_{\rm II,c}}{z_{\rm II,1}}\right)\right] & z \le z_{\rm II,c}, \\
       \exp\left[-\frac{1}{2}\left(\frac{z-z_{\rm II,c}}{z_{\rm II,2}}\right)\right] & z > z_{\rm II,c}. \\
    \end{array}\right.
\end{eqnarray} 
The parameters of these RLFs are summarized in Table. \ref{tb:rlf}. As in \citet{sta06}, the conversion factor of the cosmology $\eta(z)$ is 
\begin{equation}
\eta(z)\equiv\frac{d^2V_W/d\Omega dz}{d^2V/d\Omega dz}.
\end{equation}
The comoving volume element of cosmology in \citet{wil01} is
\begin{equation}  
\frac{d^2V_W}{d\Omega dz} = \frac{c^3z^2(2+z)^2}{4H_{0,W}^3(1+z)^3}, 
\end{equation} 
where $c$ is the speed of light and $H_{0,W}$ is $50 \rm{km/s/Mpc}$.
That of our standard cosmology is

\begin{deluxetable}{rr}[b]
\tabletypesize{\scriptsize}[h]
\tablecaption{The parameters of the RLF \label{tb:rlf}}
\tablewidth{0pt}
\tablehead{\colhead{} & \colhead{\citet{wil01}} }
\startdata
$\log_{10}(\rho_{\rm I,0}^a)$ & $-7.523$ \\
$\alpha_{\rm I}$ & 0.586\\
$\log_{10}(L_{\rm I,c}^b)$ & $26.48$\\
$z_{\rm I,c}$ & 0.710\\
$k_{\rm I}$ & 3.48\\
$\log_{10}(\rho_{\rm II,0}^a)$ & $-6.757$ \\
$\alpha_{\rm II}$ & 2.42\\
$log_{10}(L_{\rm II,c}^b)$ & $27.39$\\
$z_{\rm II,c}$ & 2.03\\
$z_{\rm II,1}$ & 0.568\\
$z_{\rm II,2}$ & 0.956
\enddata
\tablenotetext{a}{In units of ${\rm Mpc}^{-3}$.}  
\tablenotetext{b}{In units of W/Hz/Sr.}
\end{deluxetable}

\begin{equation}
\frac{d^2V}{d\Omega dz} = \frac{cd_L(z)^2}{H_0(1+z)^2\sqrt{(1-\Omega_M-\Omega_\Lambda)(1+z)^2+\Omega_M(1+z)^3+\Omega_\Lambda}},
\end{equation}
where $H_0$ is $70 \rm{km/s/Mpc}$. 

Since Equation \ref{eq:lrlg} is for radio luminosities at 5GHz in the unit of erg/s, we assume spectral index $\alpha_r=0.8$ for all radio galaxies to convert 151 MHz luminosity to 5GHz luminosity as assumed in \citet{wil01}. Although $\alpha_r$ would affect the fraction of gamma-ray loud radio galaxies in radio galaxy population, $\alpha_r$ does not affect main results on the EGRB calculation in this paper because our GLF is normalized to the cumulative source count distribution of the gamma-ray loud radio galaxies detected by {\it Fermi}.

\subsection{Source Count Distribution}
\label{subsec:lognlogs}
The normalization factor $\kappa$, which corresponds to the fraction of gamma-ray loud radio galaxies against all radio galaxies, is determined by the normalizing our GLF to the source count distribution of the {\it Fermi} radio galaxies, which is sometimes called logN--logS plot or cumulative flux distribution. Source count distribution is calculated by
\begin{equation}
N(>F_\gamma)=4\pi\int_{0}^{z_{\rm max}}dz \frac{d^2V}{d\Omega dz}\int_{L_\gamma(z,F_\gamma)}^{L_{\gamma,\rm max}}dL_\gamma\rho_\gamma(L_\gamma,z),
\end{equation}
where $L_\gamma(z,F_\gamma)$ is the gamma-ray luminosity of a blazar at redshift $z$ whose photon flux at $>$100 MeV is $F_\gamma$. Hereinafter, we assume $z_{\rm max}=5$ and $L_{\gamma,\rm max}=10^{48} {\rm erg/s}$ in this study. These assumptions hardly affect the results in this study.

Since the completeness of the {\it Fermi} sky survey depends on the photon flux and photon index of a source, we need to take into account this effect (so called the detection efficiency) to compare GLF with the cumulative source count distribution of gamma-ray loud radio galaxies. The detection efficiency of {\it Fermi} is shown in Figure 7 of \citet{abd10_marco} for the sources in the 11-month catalog with test statistics TS$>50$, at the Galactic latitude $|b|>20^\circ$, and with a mean photon index of 2.40, which is similar to that of gamma-ray loud radio galaxies (see Section \ref{sec:sam}). It is shown that results for blazar source count distribution analysis did not change even if samples with $|b|>15^\circ$ are included. Furthermore, even if they include samples with TS$>25$, the systematic uncertainties are small. Therefore, we adopt the detection efficiency shown in \citet{abd10_marco} to our samples from the {\it Fermi} 11-month catalog, although not all the samples locates $|b|>20^\circ$ or have TS$>50$.

\begin{figure}
  \begin{center}
\centering
\plotone{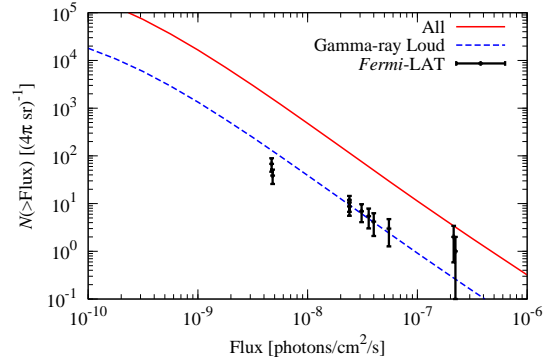}
\caption{The source count distribution of gamma-ray loud radio galaxies in the entire sky. Solid and dashed curve corresponds to the all radio galaxies and gamma-ray loud radio galaxies, respectively. The data points shows the {\it Fermi} data after the conversion of detection efficiency. The error bar shows 1$-\sigma$ statistical uncertainty.}
\label{fig:logn_logs}
\end{center}
\end{figure}

Figure \ref{fig:logn_logs} shows the source count distribution of gamma-ray loud radio galaxies. The data is after the conversion of the detection efficiency. Solid line shows in the case of $\kappa=1$ which corresponds to the case that all radio galaxies emit gamma-rays. Dashed curve corresponds to the GLF fitted to the {\it Fermi} data with $\kappa = 0.081\pm0.011$. $\sim1000$ gamma-ray loud radio galaxies are expected with 100\% complete entire sky survey above the flux threshold $F_{\gamma}(>100 {\rm MeV})=1.0\times10^{-9}\ {\rm photons\ cm^{-2}\ s^{-1}}$ above 100 MeV. We note that the current detection efficiency of {\it Fermi} at $F_{\gamma}(>100 {\rm MeV})=1.0\times10^{-9}\ {\rm photons\ cm^{-2}\ s^{-1}}$ is $\sim10^{-3}$. 
 
\section{Extragalactic Gamma-ray Background}
\label{sec:egrb}
We calculate the EGRB spectrum by integrating our GLF in the redshift and luminosity space, using the SED model shown in Section. \ref{sec:sam}. The EGRB spectrum is calculated as 
\begin{eqnarray}\nonumber
\frac{d^2F(\epsilon)}{d\epsilon d\Omega} &=&\frac{c}{4\pi}\int_0^{z_{\rm max}}dz\left|\frac{dt}{dz}\right|\int_{L_{\gamma,\rm min}}^{L_{\gamma,\rm max}}dL_\gamma \rho_\gamma(L_\gamma,z)\\ \nonumber
&&\times \frac{dL[L_\gamma,(1+z)\epsilon]}{d\epsilon}\times \{1.0 - \omega(F_\gamma[L_\gamma,z])\}\\
&&\times \exp[-\tau_{\gamma,\gamma}(\epsilon,z)],
\end{eqnarray}
where $t$ is the cosmic time and $dt/dz$ can be calculated by the Friedmann equation in the standard cosmology. The minimum gamma-ray luminosity is set at $L_{\gamma,\rm min} = 10^{39} {\ \rm erg/s}$ because there is no reported gamma-ray loud radio galaxies below this value. $\omega(F_\gamma[L_\gamma,z])$ is the detection efficiency of {\it Fermi} at the photon flux $F_\gamma$ which corresponds to the flux from a source with a gamma-ray luminosity $L_\gamma$ at redshift $z$.

\begin{figure*}
  \begin{center}
\centering
\plotone{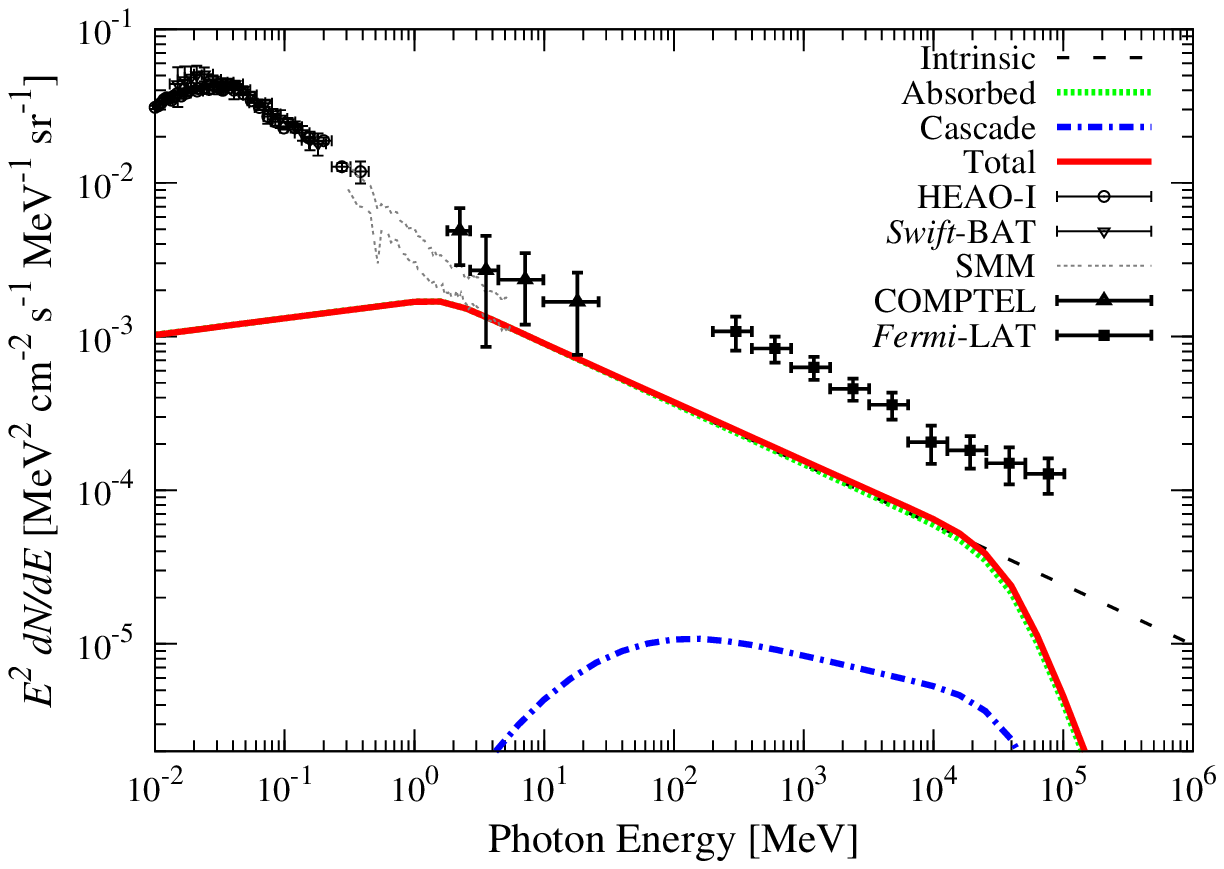}
\caption{EGRB spectrum from gamma-ray loud radio galaxies in the unit of ${\rm MeV^2cm^{-2}s^{-1}MeV^{-1}sr^{-1}}$. Dashed, dotted, dot-dashed, and solid curves show the intrinsic spectrum (no absorption), absorbed, cascade, and total (absorbed+cascade) EGRB spectrum, respectively. The observed data of {\it HEAO-I} \citep{gru99}, {\it Swift}-BAT \citep{aje08}, {\it SMM} \citep{wat97}, COMPTEL \citep{kap96}, and {\it Fermi}-LAT \citep{abd10_egrb} are also shown by the symbols indicated in the figure.}
\label{fig:egrb}
\end{center}
\end{figure*}

 High energy $\gamma$-rays ($\gtrsim$20 GeV) propagating the universe are absorbed by the interaction with the extragalactic background light (EBL), also called as cosmic optical and infrared background, \citep{sal98,tot02,kne04,ste06,maz07,rau08,fra08,raz09,gil09,fin10,kne10}. $\tau_{\gamma,\gamma}(\epsilon,z)$ is the optical depth of this background radiation. In this study, we adopt the model of \citet{fin10} for EBL and $\tau_{\gamma,\gamma}$. 
 
 The gamma-ray absorption creates electron--positron pairs. These pairs scatter the cosmic microwave background radiation to make the secondary emission component (so called cascade emission) to the absorbed primary emission \citep{aha94,wan01,dai02,raz04,and04,mur07,kne08,ino09,ven10}. We take into account the first generation of the cascade emission following the formulations as in \citet{kne08}. In the following result, the cascade emission cause a small effects on the EGRB flux. Hence, the other generations of cascade emission do not have a serious effects on our conclusion in this study.

 Fig. \ref{fig:egrb} shows the  $\nu I_\nu$ EGRB spectrum in the unit of ${\rm MeV^2cm^{-2}s^{-1} MeV^{-1}sr^{-1}}$ predicted by our GLFs. The intrinsic (the spectrum without the EBL absorption effect), absorbed, and cascade components of the EGRB spectrum and the total EGRB spectrum (absorbed+cascade) are shown. The data of HEAO-I \citep{gru99}, {\it Swift}-BAT \citep{aje08}, {\it SMM} \citep{wat97}, COMPTEL \citep{kap96}, and {\it Fermi}-LAT \citep{abd10_egrb} are also shown. As in the figure, the cascade emission does not contribute to the EGRB spectrum significantly. 

The expected EGRB photon flux above 100 MeV from gamma-ray loud radio galaxy populations is $0.26\times10^{-5}\ {\rm photons \ cm^{-2}\ s^{-1}\ sr^{-1}}$. As the unresolved {\it Fermi} EGRB flux above 100 MeV is $1.03\times10^{-5}\ {\rm photons \ cm^{-2}\ s^{-1}\ sr^{-1}}$ \citep{abd10_egrb}, the gamma-ray loud radio galaxies explains $\sim25$\% of the unresolved EGRB flux. For the comparison, recent analysis of {\it Fermi} blazars showed that blazars explains $\sim22$\% of the unresolved EGRB \citep{abd10_marco}. Therefore, we are able to explain $\sim47$\% of EGRB by radio loud AGN populations. 

To avoid the instrument dependence, we also evaluate the total EGRB photon flux (i.e. resolved + unresolved) from gamma-ray loud radio galaxies. The contribution to total EGRB is $0.27\times10^{-5}\ {\rm photons \ cm^{-2}\ s^{-1}\ sr^{-1}}$ and this corresponds to $\sim19$\% of total {\it Fermi} EGRB flux which is $1.42\times10^{-5}\ {\rm photons \ cm^{-2}\ s^{-1}\ sr^{-1}}$ \citep{abd10_egrb}.

Figure \ref{fig:egrb} also shows that gamma-ray loud radio galaxies would also contribute to the MeV EGRB at $>1$ MeV significantly. Although it has been suggested that radio quiet AGNs \citep{ino08}, MeV blazars \citep{aje09}, MeV DM annihilation \citep[e.g.][]{ahn05a}, it is still uncertain because of luck of observational evidences. As shown in this study, gamma-ray loud radio galaxies would also be a candidate for the origin of MeV background. This situation will be solved by future X-ray and MeV gamma-ray experiments such as ASTRO-H\footnote{ASTRO-H : http://astro-h.isas.jaxa.jp/index.html.en} \citep{tak10} and DUAL gamma-ray mission \citep{bog10}, respectively.

 We should examine the uncertainties in the model prediction. Since the normalization of the GLF is determined from 10 samples, there is a statistical uncertainty of 32\% in its normalization of the EGRB at 68 \% confidence level. The correlation of the radio and gamma-ray luminosities also has uncertainties in their slope and normalization as in \S.\ref{subsec:lrlg}. By taking into those uncertainties, the fraction of gamma-ray loud radio galaxies in unresolved EGRB varies from $\sim10$\% to $\sim63$\%. Furthermore, as discussed in \citet{ste10}, the energy dependence of source confusion effects would alter the EGRB spectrum below $\sim$1 GeV. However, the angular resolution is also depends on the position of the source in the field of view \citep{atw09}. The further careful evaluation is required to discuss the source confusion effects.
 
\section{Discussion}
\label{sec:dis}

\subsection{Fraction of Gamma-ray Loud Radio Galaxies}
\label{subsec:gam}
Since jet is brighter to observers with a smaller viewing angle from the jet axis because of the beaming effect, the fraction of gamma-ray loud radio galaxies $\kappa$ would be related with the viewing angle. It is believed that radio galaxies have bipolar jets \citep{urr95}. The fraction of radio galaxies with the viewing angle $<\theta$ is given as $\kappa = (1-\cos \theta)$. In this study, the fraction of gamma-ray loud radio galaxies is derived as $\kappa=0.081$ as discussed in \S.\ref{subsec:lognlogs}. Then, the expected $\theta$ is $\lesssim 24^\circ$. The viewing angle of NGC 1275, M 87, and Cen A is derived as 25$^\circ$, 10$^\circ$, and 30$^\circ$ by SED fitting \citep{abd09_ngc1275,abd09_m87,abd10_cena}, respectively. Therefore, our estimation is consistent with the observed results.

Here, beaming factor $\delta$ is defined as $\Gamma^{-1}(1-\beta\cos\theta)^{-1}$, where $\Gamma$ is the bulk lorentz factor of the jet and $\beta=\sqrt{1-1/\Gamma^2}$. If $\Gamma\sim10$ which is typical for blazars, $\delta$ becomes $\sim$1 with $\theta=24^\circ$. This value means no significant beaming effect because the observed luminosity is $\delta^4$ times brighter than that in the jet rest frame. On the other hand, if $2\lesssim\Gamma\lesssim4$, $\delta$ becomes greater than 2 with $\theta=24^\circ$ (i.e. beaming effect becomes important). \citet{ghi05} proposed the spine and layer jet emission model, in which jet is composed by a slow jet layer and a fast jet spine. The difference of $\Gamma$ between blazars and gamma-ray loud radio galaxies would be interpreted with structured jet emission model.

We note that $\kappa$ depends on $\alpha_{r}$ as in Sec. \ref{subsec:glf}. By changing $\alpha_{r}$ by 0.1 (i.e. to 0.7 or 0.9), $\kappa$  and $\theta$ changes a factor of 1.4 and 1.2, respectively. Thus, even if we change $\alpha_r$, beaming effect is not effective with $\Gamma\sim10$ but with lower $\Gamma$ value, $2\lesssim\Gamma\lesssim4$. 
 
\subsection{Uncertainty in the Spectral Modeling}
\label{subsec:unc}
As pointed in Section \ref{sec:sam}, there are uncertainties in SED modeling because of small samples such as photon index ($\Gamma$) and break photon energy ($\epsilon_{{\rm br}}$). In the case of blazars, \citet{ste96,pav08} calculated the blazar EGRB spectrum including the distribution of the photon index by assuming Gaussian distributions even with $\sim50$ samples. We performed the Kolomogorov-Smirnov (K--S) test to see the goodness of the gaussian fit to our sample and check whether the method in \citet{ste96,pav08} is applicable to our sample. The chance probability is 12\%. This means that the Gaussian distribution does not agree with the data. To investigate the distribution of photon index, more sample would be required.

\begin{figure}
  \begin{center}
\centering
\plotone{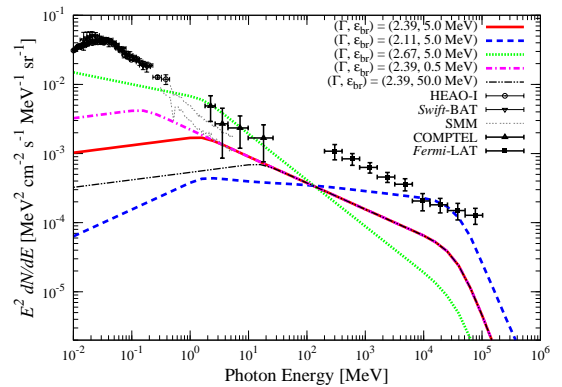}
\caption{Gamma-ray loud radio galaxy EGRB spectra in the unit of [${\rm MeV^2\ cm^{-2}\ s^{-1}\ MeV^{-1}\ sr^{-1}}$] for various SED parameters, $\Gamma$ and $\epsilon_{\rm br}$. The curves shown in this figure is the total EGRB spectrum (absorbed + cascade). Solid, dashed, dotted, dot-dashed, double-dotted dashes curves corresponds to the EGRB spectrum with $(\Gamma,\epsilon_{\rm br})=$(2.39, 5.0 MeV), (2.11, 5.0 MeV), (2.67, 5.0 MeV), (2.39, 0.5 MeV), and (2.39, 50.0 MeV), respectively. The observed data shown here are same as Figure \ref{fig:egrb}.}
\label{fig:egrb_index}
\end{center}
\end{figure}

We evaluate the uncertainties in SED models by using various SEDs. Figure \ref{fig:egrb_index} shows the total EGRB spectrum (absorbed + cascade) from the gamma-ray loud radio galaxies with various photon index and break energy parameters. The contribution to the unresolved {\it Fermi} EGRB photon flux above 100 MeV becomes 25.4\%, 25.4\%, 23.8\% for $\Gamma=$ 2.39, 2.11, and 2.67, respectively. In the case of $\Gamma=2.11$, the contribution to the EGRB flux above 10 GeV becomes significant. For the MeV background below 10 MeV, the position of the break energy and the photon index is crucial to determine the contribution of the gamma-ray loud radio galaxies. As shown in the Figure \ref{fig:egrb_index}, higher break energy and softer photon index result less contribution to the MeV background radiation. To make a further discussion on the SED modeling, the multi-wavelength spectral analysis of all GeV observed gamma-ray loud radio galaxies are required.

\subsection{Flaring Activity}
\label{subsec:fla}
It is well known that blazars are variable sources in gamma-ray \citep[see e.g.][]{abd09_3c454.3,abd10_var}. If gamma-ray loud radio galaxies are the misaligned population of blazars, they will also be variable sources. \citet{kat10} has recently reported that NGC 1275 showed a factor of $\sim2$ variation of gamma-ray flux. For other gamma-ray loud radio galaxies, such a significant variation has not observed yet \citep{abd10_core}. Therefore, it is not straightforward to model the variability of radio galaxies currently. In this paper, we used the time-averaged gamma-ray flux of gamma-ray loud radio galaxies in the {\it Fermi} catalog, which is the mean of the {\it Fermi} 1-year observation. More observational information (e.g. frequency) is required to model the gamma-ray variability of radio galaxies. Further long term {\it  Fermi}observation will be useful and future ground based imaging atmospheric Cherenkov Telescope, Cherenkov Telescope Array (CTA)\footnote{CTA: http://www.cta-observatory.org/} would be a key to understanding short period variabilities.

\subsection{Origin of the GeV EGRB}
\label{subsec:gev}
In this study, we find that the contribution of gamma-ray loud radio galaxies to the unresolved EGRB above 100 MeV is $\sim25$ \%. \citet{abd10_marco} recently showed that unresolved blazars can explain only $\sim22$ \% of the unresolved EGRB by analyzing one year catalog of the {\it Fermi} blazars. Therefore, the origin of rest $\sim$53\% of EGRB is still missing.

Various gamma-ray emitting extragalactic sources have also been discussed as the GeV EGRB origin. Those are intergalactic shocks produced by the large scale structure formation \citep{loe00,tot00,min02,kes03,gab03}, galaxies \citep{pav02,tho07,bha09_gal,mak10,fie10,ste10}, high Galactic latitude pulsars \citep{fau10,sie10}, kilo-parsec (kpc) size AGN jets \citep{sta06}, radio quiet AGNs \citep{ino08,ino09}, and GeV mass scale DM annihilation or decay \citep[see e.g.][]{jun96,ber00,uli02,oda05,and06,hor06,and07,ahn07,and09,kaw09}. 

\begin{figure}
  \begin{center}
\centering
\plotone{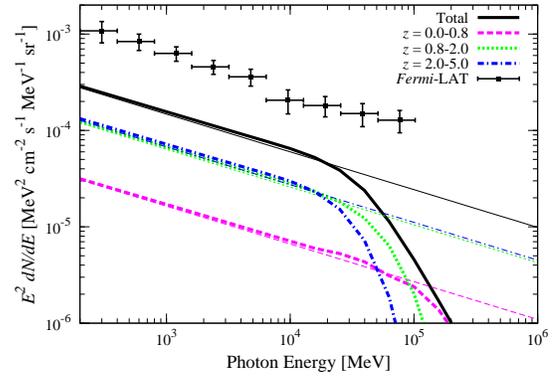}
\caption{Gamma-ray loud radio galaxy EGRB spectra in the unit of [${\rm MeV^2\ cm^{-2}\ s^{-1}\ MeV^{-1}\ sr^{-1}}$] at each redshift bins. Solid, dashed, dotted, and dot-dashed curves shows the EGRB spectrum at $z=0.0-5.0$, $z=0.0-0.8$, $z=0.8-2.0$, and $z=2.0-5.0$, respectively. The thin and thick curve corresponds to the intrinsic spectrum (not absorbed) and the total spectrum (absorbed + cascade), respectively. The observed data of {\it Fermi}-LAT \citep{abd10_egrb} is also shown.}
\label{fig:egrb_z}
\end{center}
\end{figure}

Fig. \ref{fig:egrb_z} shows the gamma-ray loud radio galaxy EGRB spectra at each redshift bins. Because of EBL, the spectrum above $>30$ GeV shows the absorbed signature. Here, the cosmological sources have their evolution peaks at $z=1\sim2$ such as cosmic star formation history and AGN activity \citep[see e.g.][]{hop06,ued03}. This means that the gamma-rays from extragalactic sources (e.g. galaxies and AGNs) will experience the EBL absorption. However, as shown in Fig. \ref{fig:egrb_z}, {\it Fermi} EGRB spectrum does not show such an absorbed signature. This might suggest that nearby gamma-ray emitting sources or sources with very hard gamma-ray spectrum would be the dominant population of EGRB above 10 GeV. To address this issue, we should await the EGRB information above 100 GeV by future observations such as {\it Fermi}. CTA would also be possible to see the EGRB at much higher energy band. We also need to examine the EBL models at high redshift. It is expected that CTA will see blazars up to $z\sim1.2$ at very high energy band $>30$ GeV \citep{ino10a}. Therefore, {\it Fermi} and CTA will be a key to understanding the origin of EGRB.

\subsection{Implication to the AGN Unification Scenario}
\label{subsec:agn}
AGN unification scenario explains various properties of AGNs in terms of viewing angle \citep{urr95}. In the scheme of AGN jet unification scenario, FRI and FRII galaxies are thought as misaligned populations of BL Lacs and FSRQs, respectively. 

From Table. \ref{tb:sam}, the mean photon index of FR Is and FRIIs are 2.27 and 2.58, respectively. Therefore, the FRI population tends to have harder spectra than the FRII population as shown in \citet{abd10_core}. This trend is also same as that between BL Lacs and FSRQs \citep{abd10_agn}. This result would support that FRIs and FRIIs are the misaligned population of BL Lacs and FSRQs.

It is also important to compare the cosmological evolutions of blazars and radio galaxies based on the recent {\it Fermi} data. Although a theoretical blazar GLF model \citep{ino09} is briefly compared with the {\it Fermi} EGRB result and the cumulative source count distribution of the {\it Fermi} blazar data \citep{ino10a,ino11_err,ino11}, comparison in redshift space has not yet been done. This is because redshifts of about a half of BL Lac samples has not yet determined \citep{abd10_agn}. 

In addition, \citet{ino09} treated blazars as one population using blazar sequence \citep{fos98,kub98}. Blazar GLF models which divide FSRQs and BL Lacs is required such as \citet{der07} to interpret the unification scenario. Since our model in this paper also does not treat FRI and FRII separately because of small samples, GLF models of gamma-ray loud radio galaxies dividing these two population are also required. 

Therefore, redshift information of all blazars and more data of gamma-ray loud radio galaxies would be required to make a comparison of the cosmological evolutions of blazars and radio galaxies.

\section{Conclusion}
\label{sec:con}
In this paper, we studied the contribution of gamma-ray loud radio galaxies to the EGRB by constructing their GLF. First, we explored the correlation between the radio and gamma-ray luminosities of gamma-ray loud radio galaxies which are recently reported by {\it Fermi} \citep{abd10_agn,abd10_core}. We found that there is a correlation $L_\gamma\propto L_{\rm 5GHz}^{1.16\pm0.02}$ by a partial correlation analysis, where $L_\gamma$ is the 0.1-10 GeV gamma-ray luminosity and $L_{\rm 5GHz}$ is the radio luminosity at 5 GHz. This slope index is similar to that of blazars.

Based on this correlation, we defined the GLF of gamma-ray loud radio galaxies using the RLF of radio galaxies. We normalized the GLF to fit to the cumulative flux distribution of {\it Fermi} samples by using {\it Fermi} detection efficiency \citep{abd10_marco}. Then, we predicted the contribution of gamma-ray loud radio galaxies to the MeV and GeV EGRB. The absorption by EBL and the reprocessed cascade emission are also taken into account. We found that gamma-ray loud radio galaxies will explain $\sim25$\% of the EGRB flux above 100 MeV and also make a significant contribution to the 1--30 MeV EGRB. Since blazars explain $\sim22$\% of EGRB, we are able to explain $\sim47$\% of EGRB by blazars and gamma-ray loud radio galaxies.

We also make an interpretation on the origin of the EGRB above 10 GeV from the point of view of the EBL absorption effect. Since the EBL absorption signature is still not appeared in the EGRB spectrum, the origin would be nearby sources or sources with hard gamma-ray spectrum. We should await the EGRB data at higher energy band for this issue.

\acknowledgments
The author thanks the hospitality of Max Planck Institute for physics at Munich where this work took place. The author also thanks M. Hayashida, D. Paneque, and H. Takami for discussion. The anonymous referee is thanked for his/her constructive suggestions. This work was supported by the Grant-in-Aid for the Global COE Program "The Next Generation of Physics, Spun from Universality and Emergence" and Scientific Research (19047003, 19740099) from the Ministry of Education, Culture, Sports, Science and Technology (MEXT) of Japan.  The author also acknowledges support by the Research Fellowship of the Japan Society for the Promotion of Science (JSPS).

\appendix 
\section{Partial Correlation Analysis}
To study the correlation between luminosities at different wavelengths, we use luminosities directly. However, the correlation in luminosity space is distorted by redshift, if samples are flux-limited. This will result in a spurious correlation. Therefore, we need to test the correlation excluding the redshift dependence. Partial correlation analysis is the analyzing method for such a condition \citep[see][for details]{pad92}. The partial correlation analysis method is as follows.

We have parameter sets of $(x_i,y_i,z_i)$, $i=1,2....,N$. Let $X_i$ be the rank of $x_i$ among the other $x$'s, $Y_i$ be the rank of $y_i$ among the other $y$'s, and $Z_i$ be the rank of $z_i$ among the other $z$'s. The Spearman rank-order correlation coefficient between $x$ and $y$ is defined to be the linear correlation coefficient of the ranks as 
\begin{equation}
r_{xy} = \frac{\Sigma_{i=1}^N(X_i - \overline{X})(Y_i - \overline{Y})}{\sqrt{\Sigma_{i=1}^N(X_i - \overline{X})^2}\sqrt{\Sigma_{i=1}^N(Y_i - \overline{Y})^2}},
\end{equation}
where $\overline{X}$ and $\overline{Y}$ is the mean of the $X$ and the $Y$, respectively \citep[see][for details]{nr92}. The correlation coefficients between $x$ and $z$, and $y$ and $z$ are also given in a same way. Then, the correlation coefficient between $x$ and $y$ excluding the dependence on the third parameter of $z$ is evaluated as 
\begin{equation}
r_{xy,z} = \frac{r_{xy}-r_{xz}r_{yz}}{\sqrt{1-r_{xz}^2}\sqrt{1-r_{yz}^2}},
\end{equation}
where $r_{xz}$ and $r_{yz}$ is the correlation coefficient between $x$ and $z$ and between $y$ and $z$, respectively \citep{ken79}. 
\end{document}